\documentclass[aps,prl,amsfonts,superscriptaddress,showpacs, twocolumn]{revtex4} 
\usepackage{epsfig}
\usepackage{bbm} 
\usepackage{multirow}
\usepackage{hyperref}
\newcommand{\eq}{\begin{eqnarray}} 
\newcommand{\en}{\end{eqnarray}}

\def\ket#1{\mathinner{|{#1}\rangle}}
\newcommand{\braket}[2]{\langle #1|#2\rangle}

\usepackage{color}
\begin{document}

\title{Zero-Transmission Law for Multiport Beam Splitters}
\author{Malte Christopher Tichy}
\affiliation{Physikalisches Institut, Albert--Ludwigs--Universit\"at Freiburg, Hermann--Herder--Strasse~3, D--79104 Freiburg, Germany}
\author{Markus Tiersch}
\affiliation{Physikalisches Institut, Albert--Ludwigs--Universit\"at Freiburg, Hermann--Herder--Strasse~3, D--79104 Freiburg, Germany}
\affiliation{Institute for Quantum Optics and Quantum Information, Austrian Academy of Sciences, Technikerstrasse 21A, A--6020 Innsbruck, Austria}
\author{Fernando de Melo} 
\affiliation{Instituut voor Theoretische Fysica, Katholieke Universiteit Leuven, Celestijnenlaan 200D, B--3001 Heverlee, Belgium}
\affiliation{Physikalisches Institut, Albert--Ludwigs--Universit\"at Freiburg, Hermann--Herder--Strasse~3, D--79104 Freiburg, Germany}
\author{Florian Mintert} 
\affiliation{Physikalisches Institut, Albert--Ludwigs--Universit\"at Freiburg, Hermann--Herder--Strasse~3, D--79104 Freiburg, Germany}
\author{Andreas Buchleitner} 
\affiliation{Physikalisches Institut, Albert--Ludwigs--Universit\"at Freiburg, Hermann--Herder--Strasse~3, D--79104 Freiburg, Germany}

\date{\today}

\begin{abstract} 
The Hong-Ou-Mandel effect is generalized to a configuration of $n$ bosons prepared in the $n$ input ports of a Bell multiport beam splitter. We derive a strict suppression law for most possible output events, consistent with a generic bosonic behavior after suitable coarse graining. 
\end{abstract}

\maketitle
The Hong-Ou-Mandel (HOM) effect \cite{Hong:1987mz} is an impressive manifestation of the bosonic quantum nature of photons. In the original experiment, two identical photons are sent simultaneously (within their coherence time) through the two input ports of a balanced beam splitter. Since no interaction between the photons takes place, one would intuitively expect the photons to propagate independently and not presume any correlations in the number of photons measured at both output ports. Surprisingly, the photons always leave the setup together, but never exit at different ports. Such coincident events at both output ports are completely suppressed. 

This effect is used in many applications: The visibility of the dip in the coincident detection rate provides a characterization for the indistinguishability of two photons \cite{Ou:2006ta,Sun:2009dk}, and therewith for the quality of photon sources. HOM setups are used to project photons onto the maximally entangled $\ket{\Psi^-}$ Bell-state, and consequently to, both, create and detect such states \cite{PhysRevLett.61.2921}. This is used for example in entanglement swapping protocols \cite{Halder:2007th} and quantum metrology \cite{Walther:2004it}. Furthermore, the nondeterministic gate operations in linear optics quantum computation \cite{Knill:2001ad} are based on the HOM effect.

It is suggestive to generalize the HOM setup for more than two photons and more than two input or output ports. Indeed, the enhancement of events with all particles in one port - bunching events - has been observed experimentally when several photons enter each of the two modes of an unbiased (i.e.,~balanced) two-port beam splitter \cite{Ou:1999rr,Niu:2009pr}. For a specially designed biased setup of three particles and three input ports, the suppression of coincident events was shown \cite{Campos:2000yf}. In the case of a Bell multiport beam splitter \cite{PhysRevA.55.2564,PhysRevA.71.013809} which redistributes $n$ incoming particles to $n$ ports in an unbiased way, it is known that coincident events are suppressed when $n$ is even \cite{Lim:2005qt}.

All these results imply important applications, from the creation and detection of multipartite qudit-entangled states \cite{Lim:2005bf,Zou:2002kb}, over the implementation of entanglement swapping protocols for many particles and the design of efficient quantum gates for qudits \cite{Scheel:2003vl}, to the experimentally controlled transition from indistinguishability to distinguishability for many identical particles \cite{Tichy:2009kl}. However, we still lack a comprehensive understanding of the $n$-particle, $n$-port generalization of the HOM effect, since the complexity of such a scattering problem scales very unfavorably with $n$: The number of interfering amplitudes as well as that of possible output events grow faster than exponentially. Hence, a detailed analysis of individual output events is prohibitive, and needs to be substituted by statistical considerations. 

This is the purpose of the present Letter, where we present a general study of the probabilities of \emph{all} possible output events of the Bell multiport beam splitter. 
Our treatment enables a general understanding of multiparticle interference effects, as well as on the average behavior of bosons. It hence unifies previous experimental and theoretical work on multiport beam splitters, and opens up new perspectives for the experimental verification and exploitation of bosonic multiparticle behavior. 

In the following we denote arrangements of $n$ particles in the $n$ modes by a vector $\vec s=(s_1, s_2, \dots s_n)$, with $s_k$ the number of particles in the output mode $k$, and  $\sum_{i=1}^n s_i = n$. For distinguishable, noninterfering particles the probability for a certain arrangement $\vec s$ reads \eq P_{\text{class}}(\vec s) = \frac {1}{n^n} \frac{n!}{\prod_{j=1}^n s_j!} . \label{classprob} \en We call this situation ``classical'', since, due to the lack of interference between the particles, probabilities are summed instead of amplitudes, and simple combinatorics applies. 
Hence, \emph{coincident events}, i.e. $\vec s_c=(1,1,\ldots 1)$, are realized with probability $n!/n^n$. \emph{Bunching events}, with all particles at one output mode $k$, correspond to $s_k=n$ and thus to $\vec s_b=(0,0,..,n,..0)$. They are realized with probability $1/n^n$ and, hence suppressed by a factor of $n!$ with respect to the coincident events. For large $n$, both events are highly unlikely, extreme cases.

The analogous problem with identical quantum particles is best formulated in second quantization. Since applications of our study are feasible with today's optical technologies  \cite{PhysRevA.55.2564}, we focus here on bosons with the following commutation relations for the creation and annihilation operators for the respective ports: \eq [\hat a_i, \hat a_j^\dagger] = \delta_{ij}\ , \quad [\hat a_i, \hat a_j]=[\hat a_i^\dagger,\hat a_j^\dagger ] =0. \en The initial state reads $ \ket{\Psi}= \prod_{i=1}^n \hat a_i^\dagger \ket 0 \label{input} $. Input port creation operators $\hat a^\dagger_i$ are mapped to output creation operators  $\hat b_i^\dagger$ via a unitary matrix $U$ \cite{PhysRevA.40.1371}, such that $ \hat b_j^\dagger = \sum_{k=1}^n U_{jk} \hat a_k^\dagger . $
Formally, the unbiased Bell multiport beam splitter under consideration corresponds to the unitary operation given by the Fourier matrix, defined for any dimension  $n$ by $ U_{jk}=e^{\frac{2 \pi i}{n} (j-1)(k-1)}/\sqrt{n} $.

The possible states with fixed particle number per port after the scattering process read  
\eq \ket{ \Phi(\vec s)} = \left( \prod_{j=1}^n \frac 1 {\sqrt{s_j!}}\left(\hat  b_j^\dagger \right)^{s_j} \right) \ket 0 \label{finstate} .\en 
In order to describe the event probability of a given arrangement $\vec s$, we define a vector $\vec d$ of length $n$ with entries that specify each particle's output port.  
It is constructed by concatenating $s_j$ times the port number $j$: \eq \vec d=\oplus_{j=1}^n \oplus_{k=1}^{s_j} (j) , \label{drepres} \en  e.g., for the arrangement $\vec s=(2,1,0,2,0)$, we find $\vec d(\vec s)=(1,1,2,4,4)$. Therewith we can write the transition probability to a specific output arrangement $\vec s$ 
\eq P_{\text{qm}}(\vec s)  = |\braket{\Psi}{\Phi(\vec s)}|^2= \frac{1}{  \prod_j s_j!  } \left| \sum_{\sigma \in P_n}  \prod_{j=1}^n  U_{d_j(\vec s),\sigma(j)} \right|^2  , \label{thebigsum}  \en
where $P_n$ denotes the set of all permutations of $\left\{1,..,n\right\}$. 
For coincident output states, we have $d_j(\vec s_c)=j$; i.e., the overlap (\ref{thebigsum}) becomes the permanent of the matrix $U$ \cite{Scheel:2004hc,Graham:1976nx}. The best known algorithm to compute the amplitude (\ref{thebigsum}) scales exponentially in $n$ \cite{Ryser:1963oa}. In information-theoretic terms, the evaluation hence remains an NP-complete problem \cite{Lim:2005bf}, despite the symmetry of $U$. 

Notwithstanding the apparent complexity of the problem, it is possible to exploit the 
symmetry of the matrix $U$ to formulate a powerful law (with its proof given in the appendix): Events characterized by $\vec s$ are \emph{strictly suppressed} if the sum of vector-components $d_l(\vec s)$ is not dividable by $n$:
\eq Q(\vec s) := \mbox{Mod}\Big( \sum_{l=1}^n d_l(\vec s) ,n \Big)  \neq 0 \ \Rightarrow \ \braket{\Psi}{\Phi(\vec s)}=0 \label{ourtheorem} .\en
Consider, e.g., $n=6$ and $\vec s_1=(2,1,2,1,0,0)$: One immediately finds $Q(\vec s_1)=2$, and this event is hence strictly suppressed. Unexpectedly though, the event $\vec s_2=(0,1,2,0,2,1)$, which is obtained from $\vec s_1$ by simple permutation, gives $Q(\vec s_2)=0$, and is actually enhanced by a factor larger than 7 as compared to the classical event probability (also see Table \ref{fullvanishingratio}). Note that the evaluation of (\ref{ourtheorem}) scales \emph{linearly} with $n$, and that -- as shown below -- our suppression law applies for \emph{most} output arrangements. It thus largely characterizes the general statistical behavior of the Bell multiport beam splitter, in a easily evaluable manner.

For a more detailed insight in the predictions of Eq.~(\ref{ourtheorem}), we first have to identify classes of final states that occur with equal probability.
In the classical case, the realization probability of any arrangement $\vec s$ remains invariant under permutation of the output ports $s_k$. Hence we can define \emph{classical} equivalence classes which identify arrangements related to each other by permutation. The amplitude (\ref{thebigsum}), however, is not invariant under arbitrary permutations of the $s_k$; i.e., two classically equivalent arrangements are not necessarily quantum mechanically equivalent. Only cyclic and anticyclic permutations leave Eqs. (\ref{thebigsum},\ref{ourtheorem})  invariant. 
This allows us to define a \emph{quantum} equivalence relation between arrangements, and the associated quantum equivalence classes. To estimate the number of suppressed arrangements that are predicted by (\ref{ourtheorem}), let us assume that the $Q(\vec s)$ are uniformly distributed in the interval $[0,\dots,n-1]$ for the ensemble of events $\vec s$. Then the probability to find a suppressed arrangement is given by the weight of nonvanishing values of $Q(\vec s)$, i.e., by $(1-n)/n=1-1/n$.
The number of equivalence classes and the number of suppressed output arrangements, shown in Table \ref{suppressedd}, for $n=2..14$, also confirm this scaling numerically.

The unsuppressed arrangements are listed in Table \ref{fullvanishingratio}, for $n=2..6$, together with their \emph{quantum enhancement}, i.e., the ratio of quantum-to-classical event probability. The suppression of coincident events for even $n$ derived in \cite{Lim:2005qt} is covered by  Eq. (\ref{ourtheorem}) as a special case: For such events, $\vec d(\vec s_c)=(1,2,3,..,n)$; hence  $\sum_j d_j(\vec s)=n(n+1)/2$ is never dividable by $n$ for even $n$. 
\begin{table*}[t]
\begin{minipage}{.3\linewidth} 
\begin{tabular}{crrrr}
$n$ & $N_{\text{class}}$  & $N_{\text{quantum}}$ & $N_{\text{law}}$ & $N_{\text{supp}}$ \\ \hline
2 & 2& 2 & 1 & 0\\ 
3 & 3&  3& 1 & 0\\ 
4&  5&   8 &  5 & 0\\ 
5&  7 &  16 & 10 & 0 \\ 
6&  11 &  50 & 38 & 2 \\ 
7&  15 &  133 & 105 & 0 \\
8&  22 &  440 & 371 & 0 \\ 
9&  30 &  1387 & 1201 & 0\\ 
10& 42&     4752& 4226 & 96\\ 
11& 56&    16159& 14575 &0\\ 
12  & 77&    56822& 51890 & 1133 	\\ 
13 & 101 &  200474 & 184626& 0 \\ 
14 & 135 &  718146 & 666114 & 2403 \\ 
\end{tabular} 
 \caption{Number of classical equivalence classes ($N_{\text{class}}$), quantum equivalence classes ($N_{\text{quantum}}$), classes that therewithin fulfill the law (\ref{ourtheorem}) ($N_{\text{law}}$), and suppressed classes which are not predicted by Eq. \ref{ourtheorem} ($N_{\text{supp}}$). }
	  	 \label{suppressedd}	 
  \end{minipage}
  \hspace{.02\linewidth}
    \begin{minipage}{.3\linewidth} 
\begin{tabular}{clc}
$n$ & $\vec s$ & Enhancement \\ \hline 
 3 & (003) & 6 \\ 
  & (111) & 3/2 \\ \hline  
 4 &   (0004) & 24 \\ 
  &   (0202) , (0121)&  8/9  \\ \hline  
$ 5$ &   (00005)  & 120 \\ 
 &   (00131),  (01103)  & 15/2 \\ 
 &  (00212),  (01022)  & 10/3 \\ 
 &  (11111)   & 5/24\\ \hline 
$ 6$ & $ (000006) $ & 720 \\ 
& (002004), (000141),  & \multirow{3}{*}{\Bigg\} 144/5 \hspace{1.2mm} } \\ 
 &(010104) ,  (000303),&  \\
 & (001032), (000222)  &\\
	   & (020202), (001113), (012021)  & 36/5\\ 
	 \end{tabular} 
 \caption{ Nonsuppressed output states, together with the corresponding quantum enhancement, i.e., the ratio of quantum to classical event probability.} 
	 \label{fullvanishingratio}
  \end{minipage}
\hspace{.0531\linewidth}
\begin{minipage}{.3\linewidth}
\includegraphics[width=5.cm,angle=0]{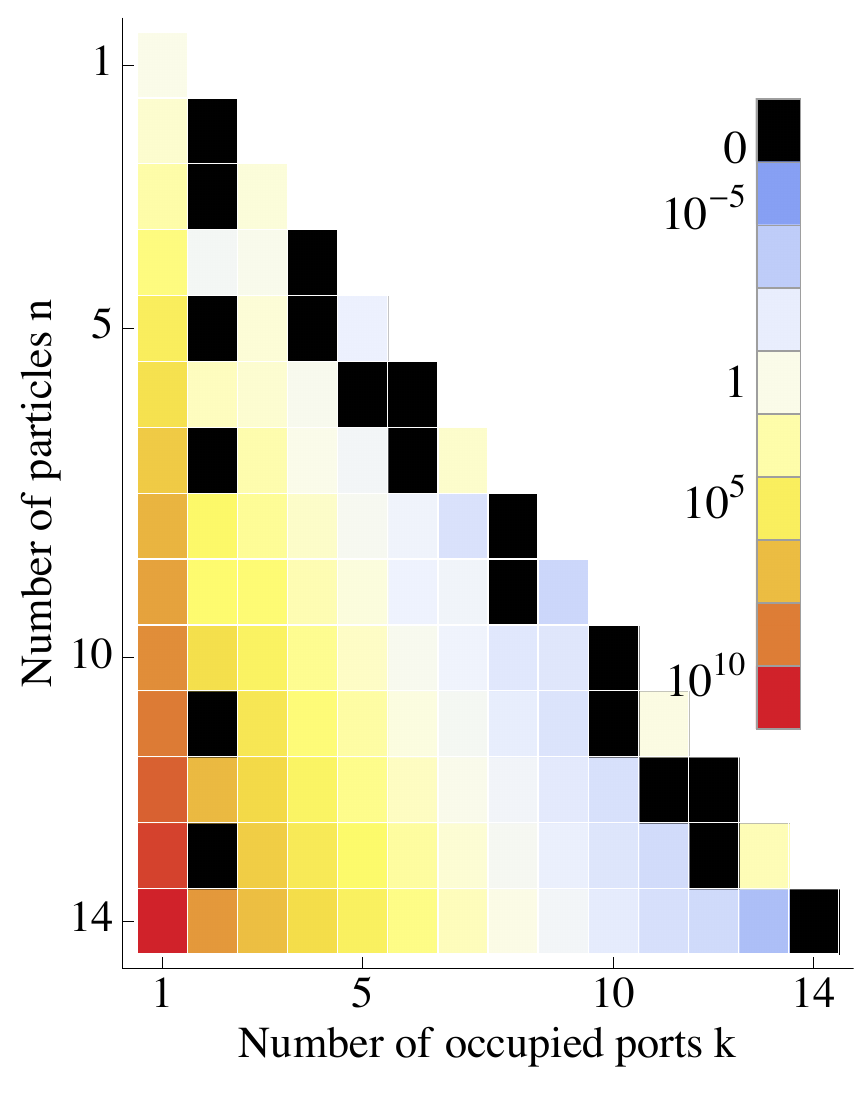} 
\caption{(color online) Quantum enhancement as a function of the number of particles (vertical axis), and of the number of occupied ports (horizontal axis).
}\label{pyramid}
\end{minipage}
\end{table*}

The implications of Eq.~(\ref{ourtheorem}) are rather counterintuitive. On the one hand, bunching events are enhanced by a factor of $n!$ with respect to the classical case, as expected due to the bosonic nature of particles: This favors states with many particles in few occupied ports. On the other hand, the number of particles in one port, or the number of occupied ports is not a direct indicator for the enhancement or the suppression of a certain event. For example, 
one intuitively expects events of the type $\vec s=(n-1,1,0,..,0)$ to be enhanced due to the bosonic nature of the particles, while they actually turn out to be strictly suppressed. Thus, at the level of the event probabilities of single arrangements, interference effects dominate, and the bosonic nature of the particles is not apparent at all. 

\begin{figure}[h]
\includegraphics[width=8.5cm,angle=0]{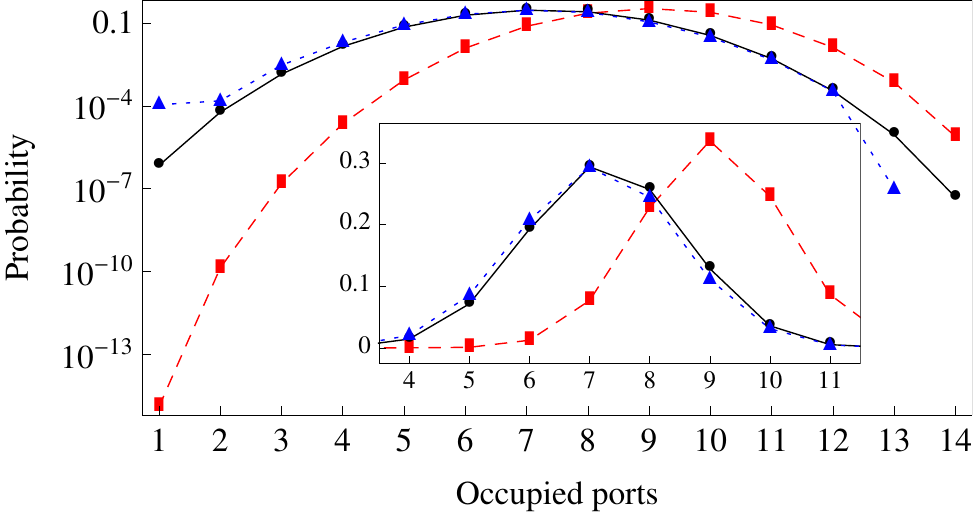} 
\caption{(color online) Event probability for a given number of occupied ports, for $n=14$. (Red) rectangles denote classical combinatorics, (blue) triangles the quantum mechanical probability distribution, and (black) circles our (bosonic) estimate for the quantum result. The inset shows the same distribution on a linear scale. Note that events with 14 occupied ports are strictly suppressed in the quantum case.}\label{nosingle2}
\end{figure}

It is, however, possible to recover a general bosonic behavior by grouping many final arrangements in larger classes which are characterized, e.g., by the number of occupied ports $k$, or by the number $m$ of particles in one port. The event probability for such a class is given by the sum of the probabilities of the single events that pertain to the class. 
Very generally, one expects that, for bosons, quantum states with large occupation numbers are  favored. This general behavior is also reflected by our formalism: According to (\ref{thebigsum}), the probabilities $P_{qm}(\vec s)$ are given in terms of a sum over permutations of \emph{scattering amplitudes}, i.e., over complex numbers of equal modulus (products of matrix-elements of $U_{jk}$).
Since these numbers typically have different phases, they tend to add up destructively.
However,
all $s_j!$ permutations $\sigma$ that interchange the $s_j$ particles that exit in port $j$ leave the scattering amplitudes invariant,
so that $s_j!$ terms in the sum have equal phases and add up constructively.
This motivates the following approximation for the transition probability (\ref{thebigsum}):
\eq P_{\text{approx}}(\vec s) = \frac{\left( \prod_j s_j! \right) P_{\text{class}}(\vec s)}{\sum_{\vec r} \left( \prod_j r_j! \right) P_{\text{class}}(\vec r)} . \label{semiesti} \en 
We show the probability distribution for the number of occupied ports, for the classical calculation (\ref{classprob}), for the bosonic quantum case (\ref{thebigsum}), and for our approximation (\ref{semiesti}), for $n=14$, in Figure \ref{nosingle2}. The distributions' expectation values correspond to the average number of occupied ports. This value can be shown to grow linearly with the number of particles $n$. However, the growth rate in the bosonic case is approx. 0.50, smaller than the value of approx.~0.63 found in the classical case. As expected, bosons always tend to occupy less output ports than in the classical case, for any $n$. Furthermore, Figure \ref{nosingle2} shows that the approximation (\ref{semiesti}) predicts the actual outcome very well for most $k$, and only fails for events with almost all or almost no sites occupied. This is easily understood, since, for very small or very large $k$, few distinct equivalence classes contribute to these event groups. 
Then, again interference dominates the event probability, rather than bosonic behavior. 
Therefore, a few general suppression effects that follow from (\ref{ourtheorem}) persist at this level, for extreme values of $k$: Events with $n-1$ occupied ports are suppressed for odd $n$, coincident events are forbidden for even $n$. If $n$ is prime, there will be never exactly two occupied ports $i,j$: The sum in (\ref{ourtheorem}) then becomes $k\cdot i+(n-k) j$ with $k$ the number of particles in port $i$. The result is never dividable by $n$ for $0<k<n$. These effects and the quantum enhancement of the classes with $k$ occupied ports are visualized in Table \ref{pyramid}. 

\begin{figure}[h]
\includegraphics[width=8.5cm,angle=0]{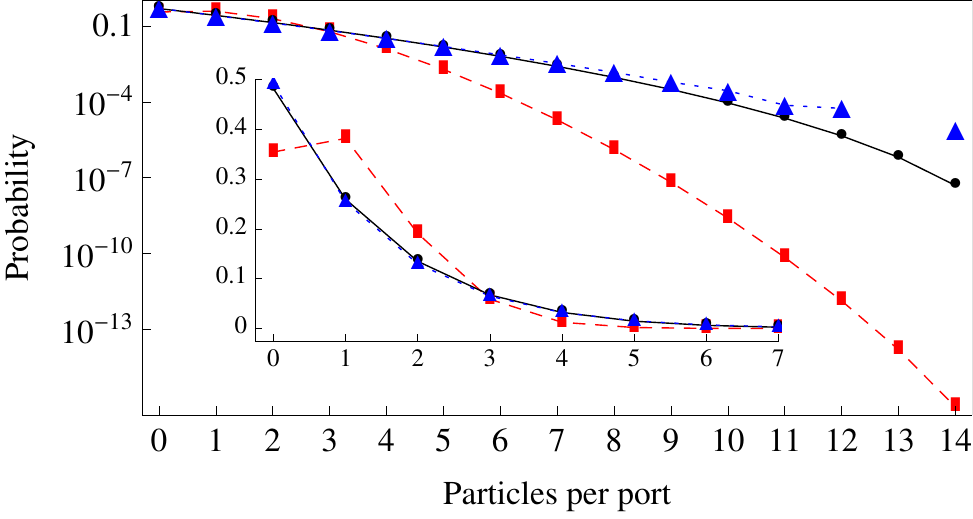} 
\caption{(color online) Probability to find exactly $k$ particles (horizontal axis) in one port, for $n=14$. (Red) rectangles denote the classical, (blue) triangles the quantum calculation, and  (black) circles the estimate (\ref{semiesti}). The inset shows the distribution on a linear scale for small $k$. Note that events with 13 particles in one port are totally suppressed in the quantum case. }\label{occudistri}
\end{figure}

Also the event probability for a given number of particles in one single port is well described by our estimate (\ref{semiesti}). 
For 14 particles, the probability distribution is shown in Figure \ref{occudistri}. Again, we see a dramatic difference between the classical and quantum case, especially for the probability to find a large number of particles in one port. 

We have generalized the HOM effect to $n$ particles and $n$ ports on two different levels: Interference effects inhibit the realization of most possible events for single transition amplitudes, while general statistical characteristics with smooth bosonic behavior emerge that are efficiently approximated by Eq. (\ref{semiesti}). On the fine as well as on the coarse grained scale, however, quantum and classical transmission probabilities differ dramatically.

In order to verify our above theoretical findings, and to statistically characterize the indistinguishability of many photons, single-photon counting detectors are required in the experiment. Since such detectors are not standard equipment yet, let us stress that also a more coarse grained measurement with bucket detectors which 
do not count the number of simultaneously arriving photons exhibits a strong quantum signature in the event statistics, as clearly spelled out in Figure \ref{nosingle2}. 

M.C.T. acknowledges financial support by Studien\-stiftung des deutschen Volkes, F.d.M. by the Belgium Interuniversity Attraction Poles Programme P6/02, and F.M. by DFG grant MI 1345/2-1, respectively.

\emph{Appendix.} Each of the $n!$ terms in the sum in Eq. (\ref{thebigsum}) can be written as an 
$n$-th root of unity. Hence, Eq. (\ref{thebigsum}) turns into $ \sum_{k=0}^{n-1} c_k e^{i \frac{2 \pi}{n} k} $, where the $c_k$ are natural numbers which give the cardinality of the following sets, defined in analogy to \cite{Graham:1976nx}, 
\eq u_r(\vec s) &=& \left\{ \sigma | \Theta_{n,\vec s}(\sigma) \equiv  \sum_{l=1}^n d(\vec s)_l \sigma(l)  = r \mbox{ mod } n\right\}, \en  with $ c_r = |u_r(\vec s)|$. 
The sum corresponds to the position of the barycenter of the set of points $\{ c_k e^{i \frac{2 \pi}{n} k} | k \in \{1,..,n \} \}$ in the complex plane. We set $Q=\text{mod}\left(\sum_{l=1}^{n} d_l(\vec s),n\right)$, and 
define an operation $\gamma$ which acts on permutations such that $\gamma(\sigma)(k) = \sigma(k) + 1 \mbox{ mod } n $. It is immediate that $ \Theta_{n, \vec s}\left( \gamma(\sigma) \right) = \Theta_{n,\vec s} \left( \sigma \right) + Q. $
Thus, if $Q\neq 0$, the repeated application of $\gamma$ gives us a bijection between all pairs of $u_{r+a \cdot Q}$, for $a \in \left\{0,1,..,n-1 \right\}$. Hence, we find \eq \forall r\in \{0,..,n-1\}, \forall a \in \mathbbm{N}: c_{r+a \cdot Q} =c_{r} .\en
Therefore, if $Q\neq 0$, the set of points $\{c_k e^{i \frac{2\pi}{n} k }| k \in \{1,..,n \}  \}$ describes several interlaced polygons centered at the origin, ensuring that the sum vanishes. On the other hand, it is also possible for the barycenter of the structure spanned by the $c_k$ to lie in the origin, even though the set of points is not described by polygons. Therefore the reverse of the law (\ref{ourtheorem}) does not hold.

\end{document}